# Towards Laser Driven Hadron Cancer Radiotherapy: A Review of Progress

K.W.D. Ledingham,
SUPA, Department of Physics, University of Strathclyde, Glasgow G40NG, Scotland
and Universıty of Selcuk, Faculty of Science, Department of Physics,
Konya, 42031,Turkey

P.R. Bolton, N. Shikazono,
Kansai Photon Science Institute, Japan Atomic Energy Agency,
8-1-7 Umemidai, Kizugawa, Kyoto 619-0215,

C-M. Ma
Radiation Oncology Department, Fox Chase Cancer Center,
7701 Borehole Ave., Philadelphia 19111, USA

## Abstract

It has been known for about sixty years that proton and heavy ion therapy is a very powerful radiation procedure for treating tumours. It has an innate ability to irradiate tumours with greater doses and spatial selectivity compared with electron and photon therapy and hence is a tissue sparing procedure. For more than twenty years powerful lasers have generated high energy beams of protons and heavy ions and hence it has been frequently speculated that lasers could be used as an alternative to RF accelerators to produce the particle beams necessary for cancer therapy. The present paper reviews the progress made towards laser driven hadron cancer therapy and what has still to be accomplished to realise its inherent enormous potential.

## Contents

**Preamble**

About 4% of people in developed countries are diagnosed with cancer each year and more than 50% of these patients receive radiation treatment, making radiotherapy along with surgery the most common as well as the most successful form of cancer therapy. Although radiotherapy is predominantly performed using photons, proton therapy is quickly gaining popularity due to its ability to irradiate tumours with higher doses and greater spatial selectivity (localization). This means that with protons (and other ions) healthy tissue can be more readily spared during irradiation than using electron or photon therapy. However, patient access to proton and heavy ion ($C^{+6}$) treatment is highly restricted due to high cost and consequent limited availability.

The use of charged hadrons in cancer therapy is almost as old as the discovery and development of the cyclotron in 1932. The first suggestion that fast protons could be used in radiology was made by "Bob" Wilson in 1946 (1) although it was not until 1954 that the first patients were treated. According to Wikipedia (2) (http://en.wikipedia.org/wiki/Proton_therapy), as of 2012 there were 39 particle therapy facilities in the world located in Canada, China, Czech Republic, France, Germany, Italy, Japan South Korea, Poland, Russia, South Africa, Sweden, Switzerland, the UK and the USA. More than 50% of the centres however are in the USA and Japan. Almost 100,000 patients have been treated with hadrons (protons and carbon ions). To treat tumours that are deeply seated within the body, proton energies above 250 MeV are needed while for carbon ions 400 MeV/amu is required.

Nonetheless, radiotherapy using accelerator based hadron beams is well-established. A selection of the many excellent publications on this subject is given in references **(3-11)**. The authors of reference **(3)** claimed that "if 200 MeV proton accelerators would be as cheap and small as the 10 MeV electron linacs used in conventional radiotherapy, at least 90% of the patients would be treated with proton beams". This indicates how effective hadron therapy is in cancer treatment. However caution must be exercised with this statement because there are very few randomized controlled trials comparing treatment with protons and photons (we further address this issue later in this work).

As stated earlier, what is hampering the wider application (and access) of this therapy mode around the world is the high cost of these centres. The authors of **(3)** estimate that the cost of establishing a conventional accelerator-based hadron therapy centre is in the range, 130-150M€ which is prohibitive for many developing countries (relevant cost comparisons will be subsequently addressed in this work).

Regarding future perspectives, the authors of **(3)** have stated that "further in the future, the first proton single room facility based on the illumination of a thin target with powerful ($10^{18}$ - $10^{20}$ W/cm$^2$) and short (30-50fs) laser pulse is expected". The basis of their statement is a shared optimism that the cost and size of laser-based proton therapy centres might be greatly reduced in the future **(5)**. Reduced cost and size will be critical for facilitating general access to hadron radiotherapy. In fact, a reasonable design constraint for the future is that full systems (novel sources, laser-based or otherwise, and other associated treatment components) should be accommodated within the infrastructure of existing hospitals (a consistent prediction has been made by Malka *et al* (see section *1.3* ))

The overall challenge for laser-driven ion beam radiotherapy is to develop well-controlled, reliable energetic ion beams of very high quality that can meet stringent medical requirements with respect to physical parameters and performance and therefore represent a viable alternative in an advancing state-of-the-art for radiotherapy.

## Introduction

The history of innovative laser applications demonstrates its ubiquitous character. The feasibility of using lasers for hadron radiotherapy has been considered for more than a decade based on energetic ion yields from intense laser-plasma experiments. It has been one of the main applications considered in laser-driven ion acceleration studies **(17-22)**. We also include in this work laser driven electron beams for radiation therapy and radiobiology **(23,24)** as well as laser driven proton beams for radiobiological studies **(25-27)**. Moreover for completeness we have included finally a section dealing with laser driven fragmentation studies as a very recent cancerous tissue diagnostic which has received much attention in the last year **(120-124)**

Laser technology and relevant laser-plasma physics as drivers are advancing rapidly. Outstanding progress has been made in high-power laser technology in the last decade with laser powers reaching the petawatt (PW) level. At present, there are fifteen PW lasers built or being built around the world while even higher power lasers (exawatt and beyond) are being considered by some. With adequate focusing petawatt lasers can generate peak electric fields of order $10^{12}$ V/cm with relatively efficient conversion to relativistic electrons with energies in excess of 1 GeV. In turn these electrons can generate beams of protons, heavy ions, neutrons and high-energy photons. Given the potential for energetic laser-driven particles we can now consider experiments in laser laboratories that are normally associated with conventional nuclear accelerators and reactors. A number of reviews describing the production of ion and electron beams by intense laser irradiation have been published during the last few years e.g. **(13-16)**.

In 2007 **(28)** Linz and Alonso published a seminal paper entitled "What will it take for laser driven proton accelerators to be applied to tumour therapy" which acknowledged the potential of a compact laser based system but warned that the optimism of laser physicists should not raise undue expectations in the medical community. Reduced cost and size are shared general aims in laser-driven particle beam development for radiotherapy. However, many scientific and technical challenges must first be met to achieve this and to confirm unique laser-driver capabilities. Reaching medically relevant particle kinetic energies is obviously essential, yet equally important is the demonstration of repetition-rated, well-controlled beamlines at these energies with suitable bunch parameters that are highly reproducible. A systems mindset is necessary to incorporate, optimize and even exploit the multiple technologies that must be combined. We can refer to the full system as the integrated laser-driven ion accelerator system or ILDIAS (ILDEAS for electrons). ILDIAS components therefore include the laser driver, the laser target, instrumentation for diagnostics and control and beam line (ion) optics. In this novel accelerator we can regard the laser plus target (which includes the induced laser-plasma at the target) as the accelerator source.

The limited scope of this work is to review global progress toward laser-driven hadron therapy specifically in terms of particle kinetic energy and other intrinsic source features (such as angular divergence and duration of the emitted particle bunch) of the laser-driven acceleration in a plasma environment.

## 1. Pulsed Ion Beams (protons and carbon ions): Laser-driven Source Features

Proton and heavy ion therapy are important and efficient sources of therapeutic irradiation on account of the capability for dose localization. **Fig. 1** illustrates the well-known relative dose of therapeutic irradiation beams as a function of depth in tissue.

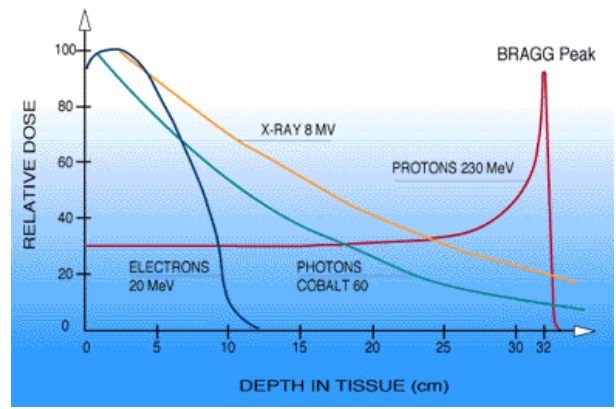

**Fig. 1** The relative radiation dose deposited as a function of tissue depth for various particle and photon beams

It can be seen that commonly used X-ray beams feature a maximum dose at a shallow tissue depth but continue to deposit energy throughout the x-ray range. On the other hand the maximum deposition of proton energy is near the end of its range (resulting in the well-known Bragg Peak phenomenon) which can coincide with tumour depth with an appropriate energy choice. Healthy tissue located upstream and downstream of the tumour is less affected by the irradiation. This localization defines the strength and effectiveness of the applied irradiation procedure. Furthermore carbon ions are biologically more effective and feature improved dose localization over protons such that healthy tissue is better spared. For example **Fig. 1** shows that, for a tumour located about 32 cm into the body, the dosage would be maximised using 200 MeV protons while the dosage to healthy tissue would be significantly reduced compared to that from an x-ray beam.

*1.1 Typical Ion Energy Spectra*

We do not discuss the laser acceleration processes in detail but refer readers to the many papers written on this subject; in particular, two new reviews (**16** and **17** Macchi, Daido). The laser and target parameter regime relevant to the Target Normal Sheath Acceleration (TNSA) scheme **(29-31)** (or a variant of it) is the most studied. A conceptual description of this process is now given. A short, high-intensity laser pulse is incident on a thin target of any number of different materials, which invariably has hydrogen-bearing contaminant layers on its surfaces. These targets are characteristically a few microns thick and because of this, the laser pulse is mostly reflected. The laser prepulse ionizes the front surface to form a preplasma of electrons and ions. The subsequent main laser pulse is partially absorbed in this preplasma and accelerates the electrons in the forward direction to relativistic velocities (i.e. high temperatures) from the front surface to the back surface. The hot electrons and cool ions at the target back surface generate a charge separation and therefore an electrostatic extraction field over a few microns which accelerates back-surface protons and other ions in this field region. The whole process occurs within a vacuum environment. A sketch of the TNSA process is given in **Fig. 2**. Extracting energetic particles from the vacuum then requires transmission through a thin vacuum window.

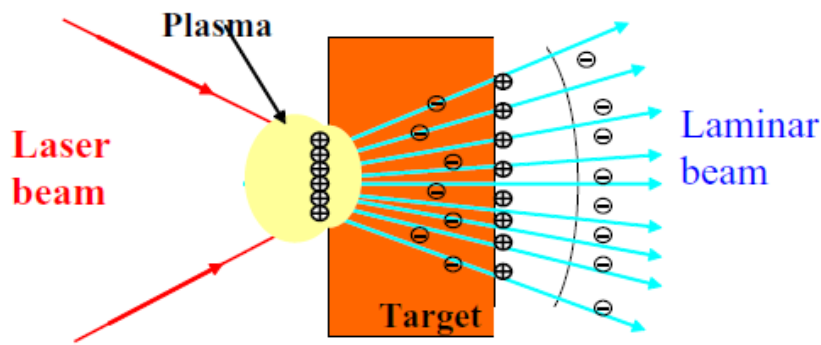

**Fig. 2** A simple sketch of ion beam generation and acceleration by the TNSA mechanism showing the extreme laminarity of the beams. At this stage the beams are not neutral although quasi-neutrality becomes evident further from the target.

Laser-accelerated ion spectra from thick targets feature a broad energy spread which is much greater than the monochromatic case typical of proton and heavy ion beams produced by conventional accelerators (in this work conventional means not laser-driven). In fact, the energy spread in the TNSA regime can exceed 100% with relatively few ions at the higher energies needed for proton irradiation of deep seated tumours. Moreover laser driven ion beams at the source (laser target) have a characteristically wide angular divergence (half angle near 10-20°) as well as a short bunch duration (ps) which are very different from conventional accelerator produced beams.

**Fig.3** shows data taken using picosecond laser pulses provided by the Vulcan laser at the Rutherford Appleton Laboratory (STFC-RAL). The displayed spectral profile is quasi-exponential with a high energy cutoff or maximum value. Laser and target (i.e. combined source) parameters can determine the maximum proton energy. We note that there are fundamentally very few particles at the cutoff energy and that the cutoff value is likely to be a demonstrably unstable (shot-to-shot) feature of the spectrum. It is not assumed to be an operation energy for a laser-driven accelerator. Instead, its value is in spectral characterization and comparison with simulated prediction. The scaling of the maximum proton energy using long pulse glass lasers (~ picosecond pulse duration) is shown in **Fig 4**. **Fig. 5** shows its variation with pulse duration and laser power where it is clearly steeper for shorter pulse durations. The angular divergence of the emitted ion beam behind the target is much smaller than that in front of the target.

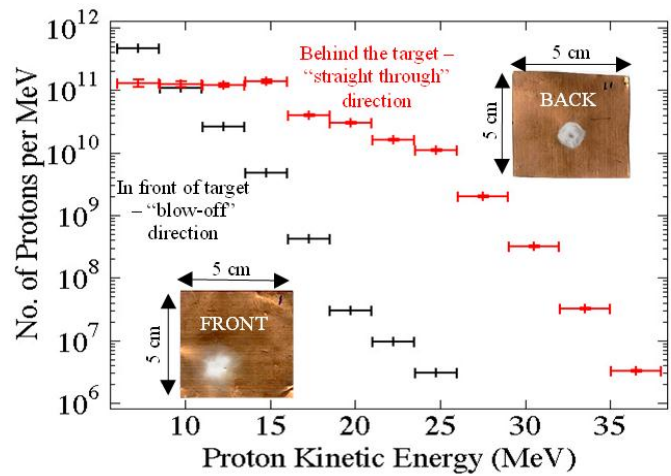

**Fig. 3** Proton energy spectrum from copper activation stacks in front of and behind the target. Inserts show the first copper piece in each stack containing proton spatial information (**32**). It should be pointed out that if radioactivity in activation stacks is used as a proton diagnostic then to be an unambiguous diagnostic for protons the nuclear reaction must be identified e.g using Cu activation stacks the $^{63}$Cu (p,n)$^{62}$Cu with a 10 minute half-life was measured. A number of authors have used just a general radioactive measurement as a diagnostic which unfortunately also includes neutron and photon activated sources of radioactivity.

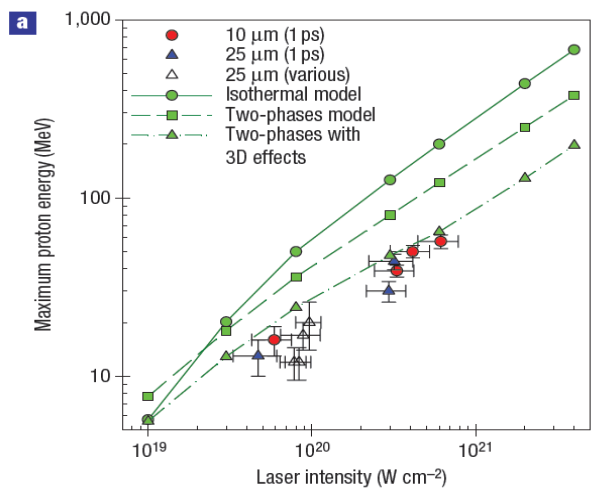

**Fig. 4** Maximum proton energy detected as a function of laser intensity obtained by variation of the laser pulse energy with a constant pulse duration of 1 ps for Al targets with thicknesses of 10 μm (red circles) and 25 μm (blue triangles). The open triangles correspond to laser shots on 25 μm targets for which the energy and pulse duration are both varied. The model data green curves is described in **(33).**

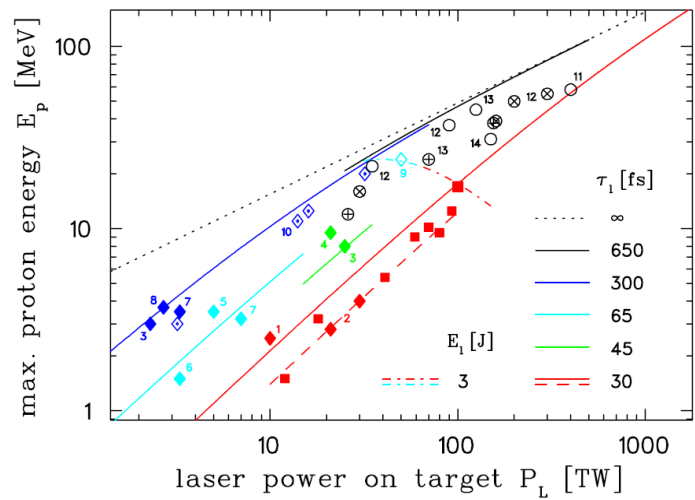

**Fig. 5** Energy scaling of the accelerated protons as a function of the laser power on target as well as pulse width. The many different contributions from various authors are given in **(34).** Reproduced by kind permission of the authors and the publishers.

For a given laser power on target the maximum proton energy increases markedly with pulse duration as revealed in Fig. 5. At the time of writing the maximum reported proton energy was about 70 MeV in the TNSA acceleration regime. This is well below the energy necessary for ion beam treatment of deep seated tumours. However, ocular tumours could be treated at this energy.

*1.2 Spectral Control (with lasers and targets)*

Most of the early work on laser acceleration of ions was conducted with thick aluminium targets (micron). Using ultrashort laser pulses (33 fsec duration) with a high contrast ratio ($10^{10}$) at the Lund Laser Centre, Neely and his team **(35)** showed that as the target thickness decreased from 30um to 100 nm the maximum proton energy and energy conversion efficiency (on target laser pulse energy to full spectrum proton kinetic energy) increased. This is shown in **Fig. 6.**

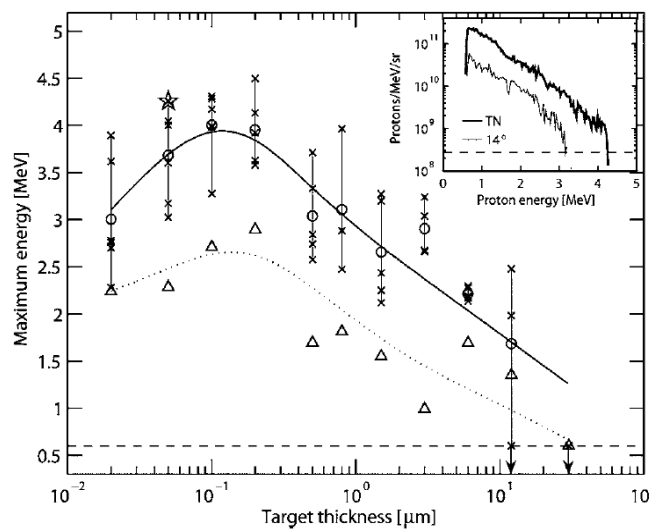

**Fig. 6** The proton spectra at $0^0$ (target normal) and $14^0$ (35) as a function of aluminium target thickness. The contrast ratio is ~$10^{10}$ and laser intensity is ~$10^{19}$ W/cm$^2$. The maximum energy increases as the thickness decreases with a maximum near 100nm. Reproduced by kind permission of the authors and publishers.

However recent developments both theoretical and experimental in laser driven ion acceleration have shown considerable promise for increasing the maximum energy efficiency of the acceleration process taking advantage of the so-called transparent overdense regime **(36-41)**. The transition to this regime occurs when the target is thin enough, typically tens to hundreds of nanometers, and the laser intensity and contrast are high enough, for all of the target electrons to be removed from the material during the laser-target interaction. Relativistic transparency enables the laser field to interact with the entire target while it is still overdense. As a result the entire ion population can be accelerated in unison. This induced dynamic transparency is a feature of both "Break-Out Afterburner" (BOA) Acceleration and Radiation Pressure Acceleration (RPA). BOA has been explored extensively over the last three years at the Los Alamos National Laboratory both experimentally and in simulations where it has been shown that ions (protons and carbon ions) of significantly increased energy can be produced. Close to 160 MeV protons and 700 MeV for fully ionised carbon ions are recorded **(41-43)**. A proton spectrum obtained in the BOA regime is shown in **Fig. 7.**

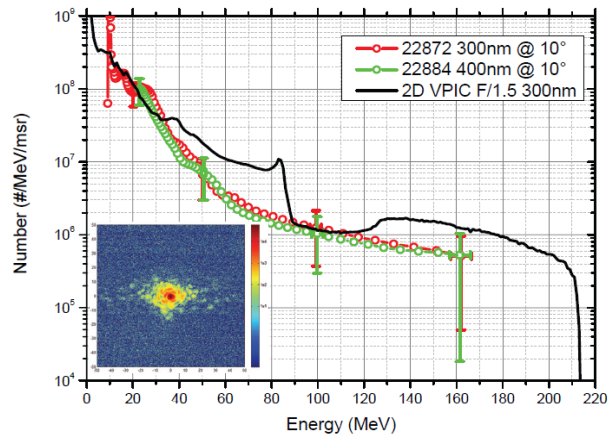

**Fig. 7** Proton spectra from the LANL Trident laser with 80J, 600fs pulses, r < 2µ spot size, target thickness 300,400nm with CH$_2$ targets. Reproduced by kind permission of the authors and publishers.

Ion acceleration by laser light pressure (Radiation Pressure Acceleration –RPA) has also received a lot of attention recently because this mechanism can potentially accelerate ions to even higher energies (than TNSA or BOA) with significantly reduced energy spread and much higher efficiency. We do not address the RPA regime in any detail and readers can consult the listed references **(44-47)**. Most of this work is theoretical and although a few experiments have been attempted significantly greater carbon ion energies must be reached in order to be clinically relevant (in this regime proton energies are reaching useful energies).

Almost all of the laser-driven proton ion beam studies have used 0.8µ wavelength photons and laser pulse durations of either 1ps or 50-100 fs. However recently very different laser parameters with another new mechanism have been applied to proton acceleration. Two groups, Haberberger *et al* **(48)** and Palmer *et al* **(49)** used an infrared $CO_2$ with a 10μ wavelength to investigate the collisionless shock-wave acceleration mechanism in a hydrogen plasma. A 60 J macropulse (which included a number of 3 ps micropulses) produced almost monoenergetic 20 MeV proton bunches (~ 1 % energy spread) with very low emittance. Simulations indicate that ~200 MeV protons could be produced with the current $CO_2$ laser technology. Moreover as laser technology improves and high power high repetition rate lasers are developed the use of a gaseous hydrogen beam target for proton acceleration can become more practical than solid targets which must be physically advanced between shots. This is shown in Fig. 8. In general however, to realize the clear rep-rate benefit typically afforded by gas sources the acceleration efficiency will need to be improved by orders of magnitude.

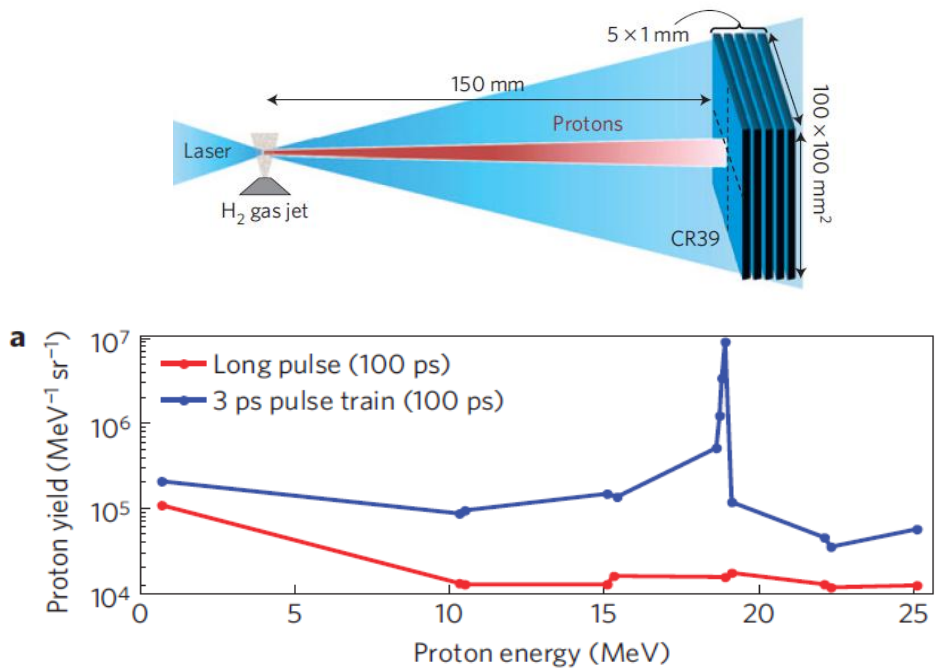

**Fig. 8** A linear polarised train of multiterawatt $CO_2$ laser pulses interacting with a gas jet. The laser with an intensity of $6.5 \times 10^{16}$ Wcm$^{-2}$ produces a monoenergetic peak at about 20 MeV. With an intensity of $7 \times 10^{19}$ Wcm$^{-2}$ computer simulations showed the peak energy increased to ~170 MeV closer to the energy needed for proton therapy. Reproduced by kind permission of the authors and publishers **(48)**.

In 2008 Keitel **(50)** and his team studied theoretically the direct high power laser acceleration of ions specifically for medical applications and found that ions accelerated by radially polarised laser have generally a smaller energy spread than those accelerated by linearly polarised lasers of the same intensity. A radially polarised beam can be focused to smaller dimensions than a linearly or circularly polarised beam but unfortunately such lasers at the required power are not yet available.

As described earlier laser-accelerated proton and heavy ion spectral profiles from thin foil targets are normally quasi-exponential. Quasi-monoenergetic spectra are desirable for ion beam radiotherapy.

Schwoerer and his team produced quasi-monoenergetic protons by irradiating microstructured targets, an idea that had been suggested by Esirkepov *et al* in 2002, **(51).** The authors pointed out that the resulting proton spectrum has a strong correlation to the spatial distribution of the protons on the target surface as shown in **Fig. 9(a).** A terawatt (TW) laser beam is focused on the front side of a metal target generating electrons, which establish a Debye sheath on the back surface as described earlier. Applying a small hydrogen-rich dot (a PMMA dot was used) on the back surface enhances the proton yield in the central part of the accelerating field, which is more homogeneous. This produces a noticeable peak in the proton spectrum as shown in **Fig. 9(b).** The figure also includes PIC simulation results for the laser and target conditions used in the experiment. An inset in the figure estimates the anticipated proton spectrum at a laser intensity of $10^{21}$ Wcm$^{-2}$.

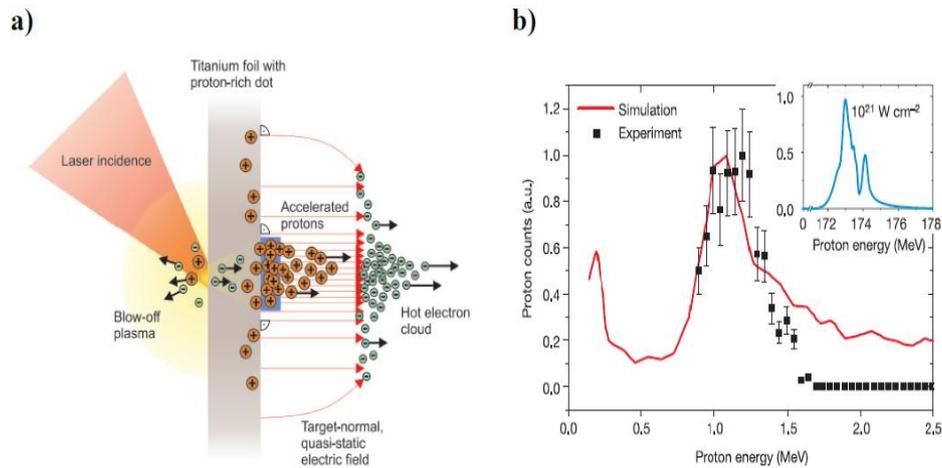

**Fig. 9** a) A laser incident on a proton dot deposited on a titanium substrate. b) Experimental peaked proton spectrum (black dots) and PIC simulation (red curve). Reproduced by kind permission of the authors and publishers **(52).** For other examples see **(53)**

A number of authors however (e.g. Brantov *et al* 2006 **(54),** Robinson and Gibbon 2007 **(55)**) pointed out this was not sufficient to explain quasi-monoenergetic features. They emphasized that a pure proton microdot target does not by itself result in a quasi-monoenergetic proton beam. Such a beam can only be produced with a very lightly doped proton target in the presence of more abundant inert and heavier ions, which generate an electrostatic shock accelerating the protons away from the surface. Their simulations suggest that beam quality in current experiments could be considerably improved by choosing microdot compositions with a 5-10 times lower proton fraction.

A laser-induced plasma optics (microlens) approach was adopted by Toncian *et al* 2006 **(56).** As shown in **Fig. 10**, two CPA laser pulses were used (CPA means chirped pulse amplification). The CPA$_1$ pulse produced a broad spectrum of protons from the foil target. Protons were directed downstream through a hollow metal cylinder irradiated by a second CPA$_2$ laser pulse. The second laser beam induces a transient focusing (radial) electric field in the hollow cylinder. For a given target-cylinder spacing the relative timing between the two laser pulses can be varied (tuned) to select a range of proton energies for focusing and transmitted energy selection.

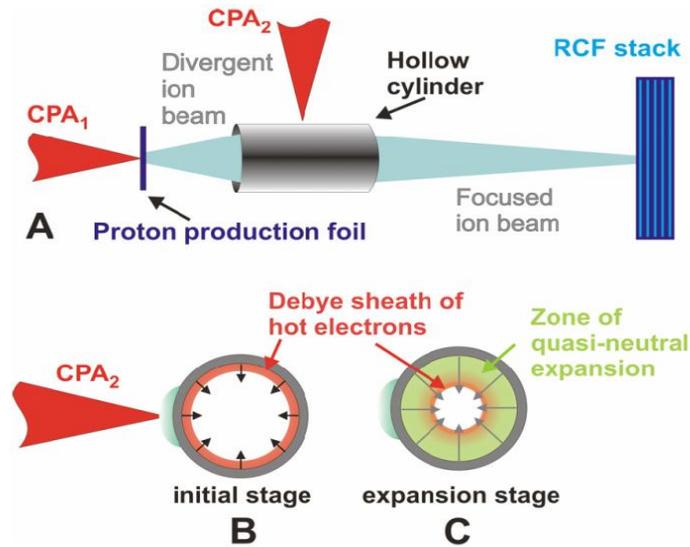

**Fig. 10.** An ultrafast laser-driven microlens to focus and energy select MeV protons. This technique is in principal scalable to high-energy protons. Reproduced by kind permission of the authors and the publishers **(56).**

This technique addresses two of the current drawbacks of laser accelerated proton bunches namely their large energy spread and the angular divergence at the source. A similar outcome could be more easily achieved using small quadrupole or pulsed solenoid magnets, an approach that is being developed by several groups including the authors of this paper **(57-59)** also showing how the proton beam can be captured and transported.  More about transport of laser produced proton beams will be discussed in section 2.

A number of other approaches to produce quasi-monoenergetic heavy ions have been reviewed previously e.g. monoenergetic deuterons from heavy water droplets **(15)**.

We now discuss very different target types that can be considered for achieving higher proton energies and proton focusing.

Zigler, Botton and co-workers have recently shown **(60)** that enhanced proton acceleration can be produced with structured dynamic plasma targets.  These targets were engineered by depositing snow on sapphire substrates which mimicked the bristles on a brush. In this case the acceleration regime was not typically TNSA **(Fig. 11)** The laser interacted with one or more bristles to provide proton energies greater than that anticipated based on normal TNSA with thin film targets.  This behaviour is shown in **Fig. 12.** At laser powers greater than 1 PW the extrapolated simulation predicts proton energies near 200 MeV.

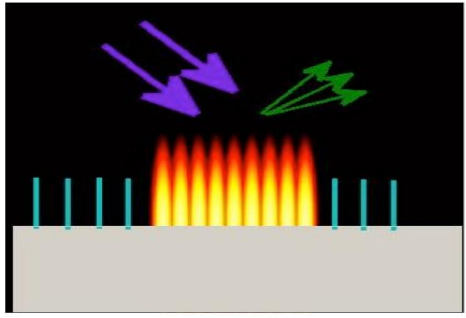

**Fig. 11** Cartoon indicating how a laser beam interacts with snow pillars about 100μ in size. The laser may interact with one or more pillars (purple arrows) and the protons come off (green arrows) from the target on the same side from which the laser approaches the target. This is not a TNSA process (reproduced by kind permission of M.Botton)

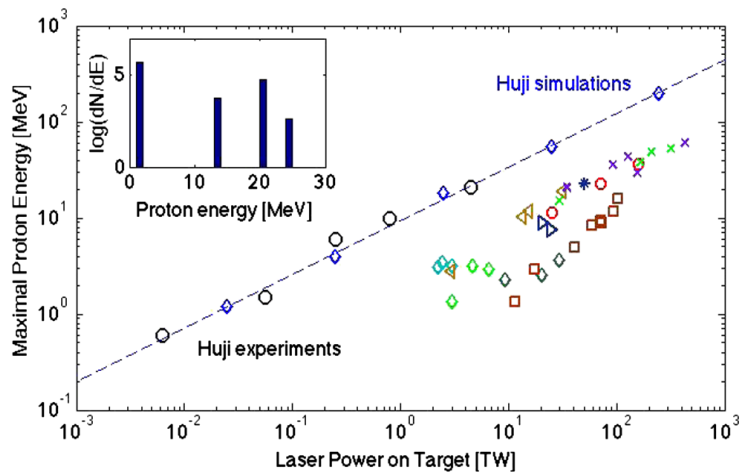

**Fig. 12** The dotted line shows the energy scaling of the accelerated protons as a function of the laser power on a snow target. All the other points below the line show the previous experimental results given in **(60)**.

It should be emphasised the novel work of Zigler and his team is based on an acceleration mechanism that is very different from normal TNSA acceleration. 'Targetry' is a rapidly advancing and critical part of ILDIAS source technology.

Recent results have been published based on simulations of laser-irradiated two-layer targets with a linearly polarised laser pulse. Such targets have an outer foam layer added to a thin solid foil. 2D PIC simulations predict 200 MeV proton yields using a realistic Gaussian laser pulse focussed to an intensity of $6 \times 10^{20}$ W/cm$^2$. **(61).**

## 1.3 Angular Divergence Control (target as an ion optic)

Quasi-exponential spectral profiles and large angular divergence represent significant technical challenges in laser-driven medical accelerator development. Mitigating beam line development issues attributed to the intrinsically large angular divergence and energy spread might require a combination of innovative ion collection optics with laser-plasma engineering.

Various methods have been used to reduce the angular divergence. Patel *et al* **(62)** and Snavely *et al* **(63)** were the first to experimentally demonstrate the ballistic focusing capability of curved target surfaces.

In 2D PIC simulations H.Y.Wang *et al* **(64)** showed that 'bulged' nanometer scale targets could generate higher energy protons (above 124 MeV) which is a factor of about two greater than TNSA results from thick foil targets. This is shown in **Fig.13.**

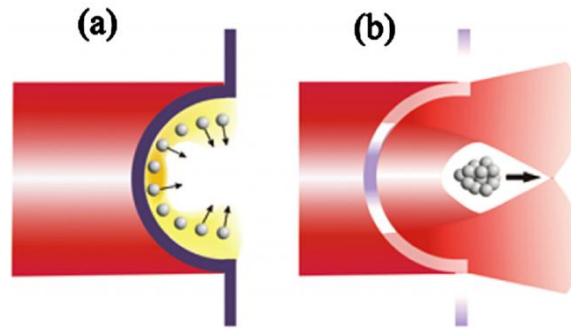

**Fig. 13** a) is conventional TNSA using a bulged target irradiated by a laser. Protons are grey balls and the electrons are the yellow cloud. b) indicates a focussed proton beam. Reproduced by kind permission of the authors and publishers.

This simulation indicates that with a laser intensity of $1.3 \times 10^{20}$ Wcm$^{-2}$ the maximum proton energy is more than 100 MeV (with a typically broad spectrum) with a conversion efficiency of 24%. If the proton energy scales as the laser intensity $I^{1/2}$ then we could expect a proton spectrum with a maximum proton energy of ~250MeV at a laser intensity of order $10^{21}$Wcm$^{-2}$.

Early experimental work by J.H.Bin *et al* **(65)** with nm diamond like carbon foils at $8 \times 10^{19}$ Wcm$^{-2}$ laser intensity have produced 5 MeV proton beams with a divergence as low as 2º. This is about an order of magnitude below divergences typical of µm targets and can therefore enhance proton fluences by about two orders of magnitude.

Generation and focussing of accelerated protons with a cone-shaped or cylindrically shaped target (downstream of the initial curved surface which provides the protons) T.Bartal *et al* **(66)** has also been demonstrated. In this sophisticated target assembly a hot electron sheath establishes a dynamic radial electric field that focuses protons emerging from the curved foil target. In a proof-of-concept demonstrated channeling of protons through the cone tip and spot size reduction have been observed in a geometry relevant to fast ignition for inertial fusion

## 2. Beam Transport and Delivery Considerations

Laser production of laser beams has yet to reach the energies and the many other qualities needed for deep seated tumour treatment. Nonetheless proton and carbon ion delivery systems have been considered for laser-driven ion beam radiotherapy (see 1998 work entitled "Prospect for Compact Medical Laser Accelerators" **(67)**). For example, **Fig. 14** illustrates a comparison between a conventional and optical gantry concept that has been suggested for the laser-driven case **(68,69)**.

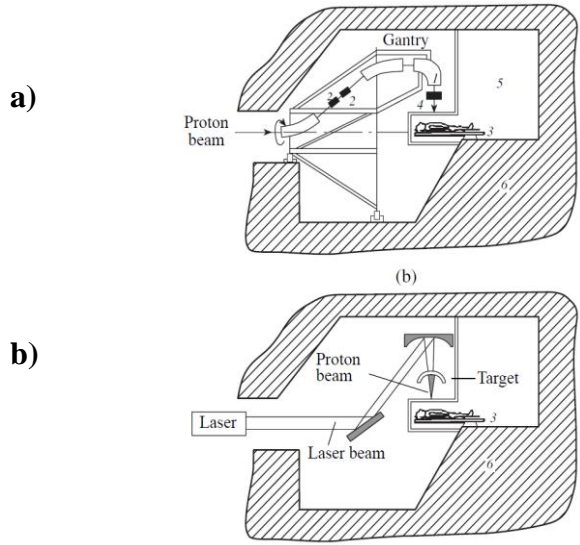

**Fig. 14** a) Conventional gantry system to direct the ion beam to the patient. This can normally be about 100 tons in weight and very expensive. b) The design of an optical (laser) gantry system which uses mirrors to convey the beam closer to the patient. Reproduced by kind permission of the authors **(68)** and publishers

One of the most influential teams involved in developing laser-driven ion beam radiotherapy is in the Radiation Oncology Department at the Fox Chase Cancer Center, in Philadelphia. In a seminal paper in 2003 and updated in 2005 **(71,72)** they presented calculations for the design of a spectrometer that allows for adjustable proton energy. A mid-plane aperture could be positioned transversely to 'tune' proton energy to a desired therapeutic energy window (see **Fig. 15**). The design accommodates the typical broad spectrum and large angular divergence of laser-accelerated protons. Furthermore, suitable upstream collection and collimation optics can improve the overall performance.

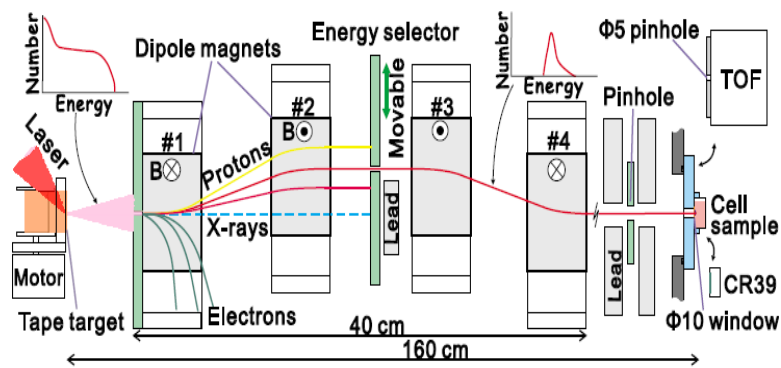

**Fig. 15** A particle selector system originally designed by Fourkal et al **(71)** This particular diagram of similar design is presented by kind permission of Yogo *et al* **(25)**. In other variants of this system further focussing quadrupoles can be placed at the output. Reproduced by kind permission of the authors and the publishers.

In a conventional accelerator driven hadron therapy unit the costs involved are so high that to make costs economical a number of different treatment rooms are included as shown in a section 7 figure. Extensive radiation shielding is required to encapsulate the accelerator, the f beam transfer section, the gantry systems and the multiple treatment rooms. The optical gantry concept (**Fig. 14**) suggests that the source (just the laser target in this case), the smaller, lighter gantry system and the patient could possibly be located in closer proximity. This can potentially reduce radiation shielding requirements and indeed the laser acceleration section requires relatively little.

The shielding design for a laser–accelerated proton therapy system has been assessed by Fan et al **(74)**. Monte Carlo simulations determined the shielding necessary for 300 MeV protons and 270 MeV electrons at a laser intensity of $2 \times 10^{21}$ Wcm$^{-2}$. Neutron production is a key problem and a layer of polyethylene enclosing the whole particle selection and collimation device was used to shield the neutrons. Encapsulating everything a layer of lead was used to reduce the photon dose from neutron capture and electron bremsstrahlung. Simulations showed that a two-layer shielding design with 10-12 cm thick polyethylene and 4 cm thick lead can effectively absorb the unwanted particles to meet the necessary shielding requirements for clinical applications.

A report describing all the elements for the development of a laser driven proton accelerator for cancer therapy was published in 2006 **(75)**. This paper discussed the treatment head of the therapy system, the specifications of the laser system, the particle selection and beam collimation system and an aperture based treatment optimisation for laser proton therapy. Other work by Ma et al **(76)** compares three novel approaches (aimed at cost effectiveness) for particle radiotherapy: dielectric wall accelerator based therapy systems, superconducting accelerator therapy systems and laser proton accelerator therapy systems.

Other new developments in the field of laser driven proton therapy have been discussed recently by Hofmann et al **(77)**. The work concentrates on collection and focusing of laser produced particles using simulated results in the high intensity Radiation Pressure Acceleration (RPA) regime. They present a scaling law for the "chromatic emittance" of the collector and apply it to the particle energy and angular spectra of the simulation output. The collector in this case is a solenoid magnetic lens. For a 10 Hz laser system, the authors find that particle collection by a solenoid magnet satisfactorily provides the requirements of intensity and emittance as needed for depth scanning irradiation in a clinical situation. The authors conclude that without ion focussing in the laser case such clinical beam intensities cannot be provided. The importance of this paper is that it addresses the RPA mechanism which is not yet commonly studied but does provide insight for the near future.

A more detailed subsequent paper of RPA simulations was presented by Hoffman *et al* **(78)**. This paper concentrates on how a beam of about $10^{10}$ protons per pulse can be used to irradiate tumours. A controlled >2Gy dose on a small tumour can be achieved by a small number of laser pulses in seconds. Spot scanning for conformal tumour irradiation requires laser pulse repetition rates exceeding 10 Hz. The conclusions of this simulation show that a combination of an RPA model, solenoid, energy selection aperture and scatterer can result in a well characterised beam suitable for therapy applications.

Finally a number of authors (e.g. Schell and Wilkens **(79)**) have cautioned that much research still needs to be done in various areas in order to implement laser driven therapy in a clinical environment and that this can take many years. They base this on the need for increasing proton energy, reducing the broad energy spread (spectral width) of protons, possible limits in the pulse repetition rate and the number of accelerated particles per pulse. Hofmann, Schell and Wilkins **(80)** do however point out that the short pulse nature of the laser processes may be ideal tools for motion adaption during radiotherapy.

## 3. Laser Driven Electron Beams for Radiotherapy

Electron beams have been used to treat cancer for more than 50 years and a review of electron beam therapy physics was carried out in 2006 **(81)**. Most electron beams generated clinically are in the 5-20 MeV range and, due to the more limited penetration depth, such beams are normally used for treatment of superficial tumours.

However electron beams above 150 MeV can readily be generated with laser irradiation of gas targets where much less laser intensity is required compared to the case for ion acceleration. They have a range in tissue of about 40 cm and hence can be used to treat deep seated tumours.

Laser-driven electron beams typically feature a quasi-monoenergetic spectral profile and are better directed than laser-driven ions (less lateral spread **(82)**). Normally the laser is directed through a supersonic gas jet and the electron spectrum is measured by a magnetic spectrometer. A typical electron spectrum is shown in **Fig. 16.**

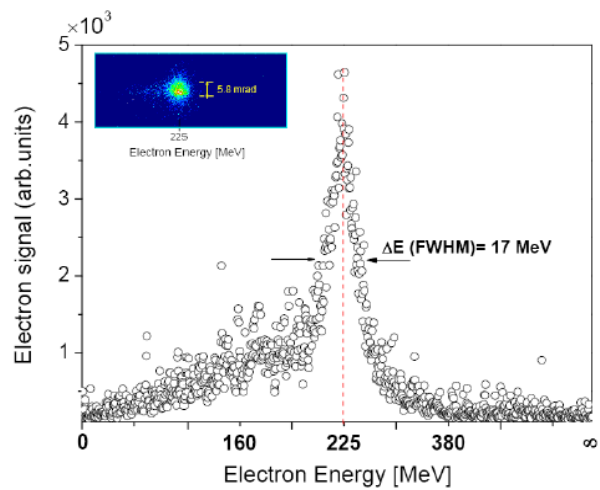

**Fig.16** Quasi-monoenergetic 225 MeV electron beam with a width of 17 MeV generated in a 4mm long gas jet at a density of $7 \times 10^{18}$ cm$^{-3}$ when irradiated with a 27 TW laser pulse **(83)**. Reproduced by kind permission of the authors and publishers.

Although most effort exploring radiotherapy of laser-driven energetic particles have concentrated on proton acceleration, Victor Malka and his group have considered laser-driven electron beams for this application. They have published extensively in this area researching a) radiotherapy with laser-plasma accelerators **(84)** both experimentally and using Monte Carlo simulation of deposited dose and b) measuring high energy electron induced damage in human carcinoma cells with electron bunches of ultrashort duration **(85)**. Treatment planning for laser accelerated very high energy electrons (now known as VHEE) has been carried out for prostate cancer and it has been found that the target coverage is better than that with a clinically approved 6 MV photon plan **(86)**.

Finally, **Fig. 17** illustrates a comparison made between a measured dose distribution from a laser-driven 120 MeV electron beam in a phantom target and that determined in Monte Carlo simulation **(87)**.

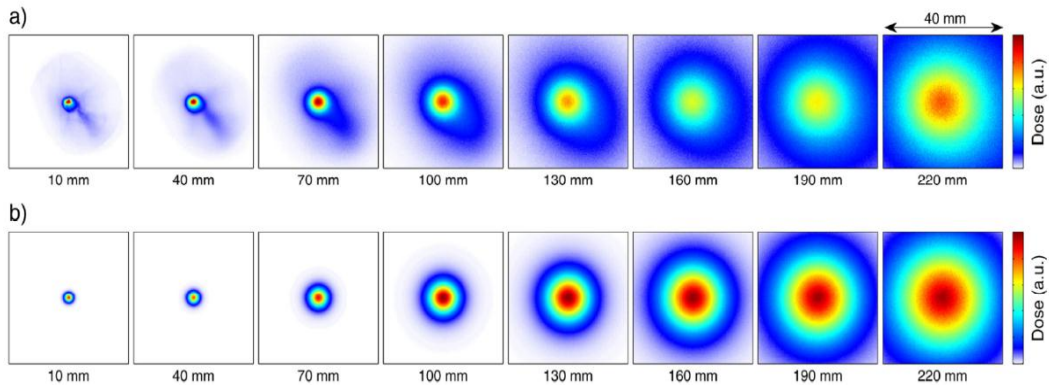

**Fig. 17** Measured a) and simulated b) two dimensional dose distributions in planes perpendicular to the beam central axis at various depths in a phantom. As the electron beam propagates through the phantom multiple Coulomb scattering occurs which deflects the electrons transversely. Published by kind permission of the authors and publishers.

The authors conclude from this work that laser-driven electron beams are quasi monoenergetic with low divergence and potentially are of interest for electron beam therapy of deep seated tumours.. The laser is more compact and delivers very high instantaneous doses within a time interval shorter than many chemical reactions. However the radiobiological effectiveness of these ultrahigh electron dose rates remains unknown.

## 4. Cellular basis of laser-driven radiation therapy

Here, we address the radiobiological basis of radiation therapy. It is useful to know whether a certain type of radiation from a certain type of source is advantageous over another for tumor curability. Curability is not determined solely by the effectiveness of radiation to individual cells, as it is well known that there are other factors such as cell to cell communication and complex interactions between tissues that could affect the sensitivity of tissues or organs **(88,89)**. However, knowledge from cellular radiobiology is at least an essential basis for assessing the effectiveness of radiation therapy. It is important to note that the biological effectiveness is not the only factor that determines the effectiveness of radiation therapy as a whole, since the energetic properties, reproducibility of doses among fractionated exposure, feasibility of irradiation etc, also affect the overall effectiveness.

Since this is a review of progress toward laser-driven radiotherapy it is also useful to know if and how the biological effect of particle radiation (radiobiological effectiveness, RBE) generated by lasers differs from that provided by conventional accelerators. This section describes basic and important issues concerning the cellular basis of laser-driven radiotherapy: (1) essential parameters relevant to cellular radiation effects, (2) the DNA damage sequence, (3) the temporal effects of high- and low-LET (linear energy transfer) radiation as it relates to the dose rate issue and (4) the oxygen effect. We also address the potential use of laser-driven radiation in radiobiological studies.

Irradiation of certain regions within a cell, normally carried out using a sophisticated microbeam apparatus, could also affect the impact of irradiating the cell. However, we do not address microbeam effects in this work. As with therapy, one usually irradiates the entire cell population of a tumor as opposed to a specific cell or region of a cell.

*4.1 Important factors for determining cellular radiation effect*

Two main quantities that determine the biological effects induced by radiation are dose and linear energy transfer (LET). The dose to a cell is the average amount of energy deposited per unit mass within the cell. Extensive work has been carried out to determine the cellular dose-response relationship for lethality, chromosome aberration, mutation induction, etc.**(90)**. Typical survival-dose response curves are shown in **Fig. 18A**. LET is the average amount of energy deposited along a unit length of the track forged by a radiation particle such as a proton. The relative biological effectiveness (RBE) is the comparison of the biological effectiveness of one type of ionizing radiation relative to another. As such, it is the ratio of doses (of two different types of radiation) required to achieve a given biological effect. RBE is usually plotted against LET **(Fig.18B)** with gamma-rays or X-rays being the standard radiation type against which other radiation types are compared. RBE is observed to slowly vary with LET over an extended range. Nonetheless, it is typically at maximum levels in the high LET range, 100-200 keV/μm which is attributed to an increased "clustering" of lesions **(91,92)**. LET is the most frequently used quantity to describe the quality of a radiation type **(90)**.

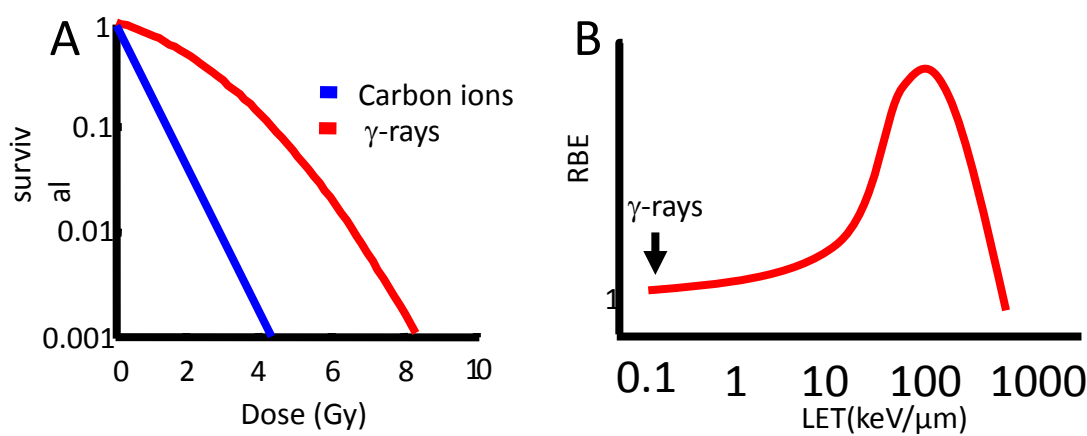

**Fig. 18**. Typical dose response curves of survival of cultured mammalian cells exposed to γ-rays and ion particles with LET of ~100keV/μm (A). Typical RBE-LET relationship demonstrating that RBE peaks near 100-200 keV/μm (B).

There are significant temporal, as well as spatial, differences between high-LET and low-LET radiation. A substantial amount of energy is deposited in a cell by an individual high-LET particle track within several tens of femtoseconds. In contrast, individual low-LET radiation tracks deposit much less energy and it is distributed over the whole cell. In the low-LET case, energy is typically delivered over much longer timescales (usually seconds or minutes using a conventional radiation source) during which chemical, biochemical and biological processes occur. In general, spatially proximal ionization and excitation events and subsequent radical formation induced by a single radiation track are generally considered to be responsible for the higher RBE of high-LET radiation. There is growing evidence that closely localized DNA lesions (clustered DNA damage) are highly relevant to the ultimate biological consequences **(91,92)**. It is well-known that the dose-depth profile significantly differs between radiation types (**Fig. 1**). Because of their higher LET, carbon ions have an advantage over protons for inactivating tumor cells. Furthermore, carbon ions are also superior in terms of dose localization

*4.2 DNA Damage Sequence and Biological Effects*

The time scale of sequential events for an isolated cell along a single particle (damage) track that lead to biological effects after irradiation **(93) (Fig. 19)** spans about twenty-one orders of magnitude. Because a cell comprises mostly water (approximately 70 %), we focus on water radiolysis.

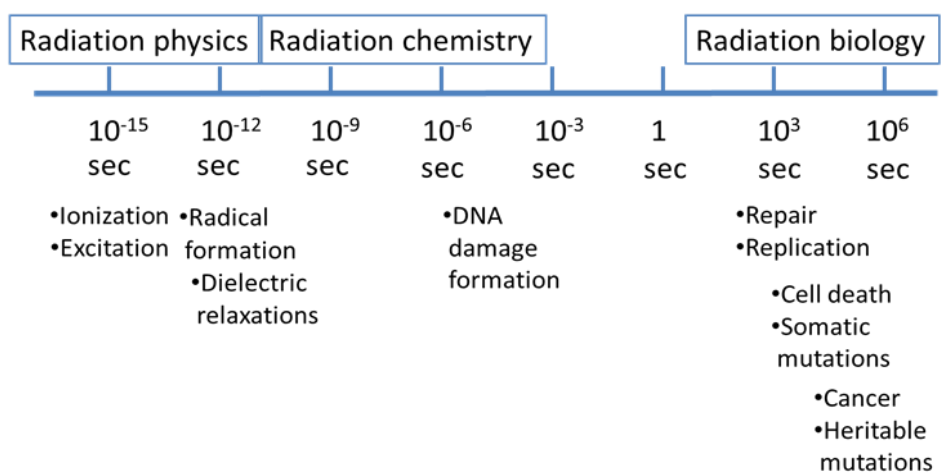

**Fig. 19.** Time scales of major events that lead to biological effects of ionizing radiation. Above the scale bar are shown the corresponding research fields that investigate the events.

The radiation effect to DNA is also considered, as it is the main target for cellular radiobiological effects **(90)** even though it constitutes only ~0.05% of the cell weight. After photons or particles pass through a cell, ionizations and excitations occur followed by dielectric relaxations. Following this initial physical stage, which occurs on the picosecond timescale, ions and ejected electrons will proceed to generate various radicals. In addition to direct ionization of DNA and DNA radicals, attacks by water-derived radicals mainly contribute to DNA damage within the picosecond to millisecond time frame. Subsequently, due to the induced DNA damage, biological events, such as DNA repair and cell cycle retardation can occur. If the damage is unrepaired and inhibits replication or cell division, cells will eventually die. Furthermore, a cell has a suicide pathway called apoptosis, in which a cell kills itself when it recognizes that the extent of damage exceeds its capacity of repair. If replication occurs with some damage unrepaired alterations of the DNA sequence (mutations) can occur. These biological events usually take place within minutes to hours. Induced mutations could become the origin of cancer. If the mutation is fixed in the cells from which egg or sperm cells are derived (germ line cells), it will lead to inherited mutation which is observed several decades after cells have been exposed to radiation.

*4.3    Dose Rate Picture*

The most notable difference between particle irradiation from laser-driven and conventional sources is the dose rate. Dose rates provided by a conventional accelerator are usually below 1 Gy/sec (but can be as high as $\sim 10^3$ Gy/sec for the spot scanning mode). At dose rates below 1 Gy/min, where the "low dose rate effect" is observed, the total irradiation time is in the range of minutes to hours. Because the irradiation time scale is comparable to that of DNA repair, the low dose rate effect is associated with repair. Previous studies have established that cellular radiosensitivity is reduced when cells are irradiated at such low dose rates.

By contrast, with laser-driven radiation, the dose rate can exceed $10^9$ Gy/sec and the irradiation time can be much shorter (typically nanoseconds). Relatively fewer studies have been conducted on ultra-high dose rates. One can assume that if an ultra-high dose rate effect exists it must be associated with specific early time physical interactions such as ionizations, excitations, and subsequently generated radicals.

We attribute an ultra-high dose rate effect to events generated by neighboring tracks that are coincident with each other. Details of this coincidence are now explained. In the low-LET case (x-rays or fast electrons) the mean energy loss per collision event is approximately 48-64 electron volts, eV (we assume 50 eV as an average value) (**93**). The absorbed energy is converted to secondary electrons, ions and radicals. The small volume of ionized water and secondary electrons along the particle track is called a "spur". The entire radiation track from a single particle is then a random succession of spurs. An ionized water molecule ($H_2O^+$) quickly decomposes to $H^+_{aq}$ and an OH radical which is mainly responsible for the DNA damage. $H^+_{aq}$ is almost motionless (diffusion distance $\sim$ 1 nm), the secondary electron range is several nm and the OH radical average diffusion distance is about 6 nm (and ultimately scavenged by cellular components)(**94,95**). The key issue is the proximity of a spur to DNA. DNA damage is correlated to a local high density of OH radicals in the volume of a nearby spur which can expand to as much as $5 \times 10^2$ nm$^3$ according to their diffusion range. For an isolated cell, the coincidence responsible for an ultra-high dose rate effect (if it exists) is therefore the coincidence of spurs. In this picture, the temporal part of this coincidence (simultaneity) is limited by the spur lifetime (of order nanoseconds). The spatial part is limited to approximately twice the spur volume (of order $10^3$ nm$^3$). In the low-LET cases, a spur coincidence would involve multiple tracks within the short lifetime of a spur. For a lethal dose such as 10 Gy (for which typical mammalian survival is less than 1 %) the spur density is about $10^{-6}$ nm$^{-3}$. Because the single spur volume is also very low, an ultra-high dose rate effect attributed to a coincidence of spurs is unlikely. This conclusion is supported by laser-driven x-ray experiments **(96-98)**, simulations based on precise physical/chemical parameters **(99)** and other experimental results **(73)**. Consequently, for an isolated cell, the effect of laser-driven radiation at an ultra-high dose rate does not likely differ from that of radiation at a moderate or a low dose rate generated by conventional accelerators.

In the high-LET case (i.e. heavier ions such as $C^{6+}$), we can consider the coincidence of spurs along a single track where significantly more energy is deposited. The single track picture makes simultaneity of spurs more likely so that ultra-high dose rate effects in radiobiology, if they exist, are more likely to occur in response to high-LET irradiation. In fact, the higher RBE of high-LET radiation could be due in part to this temporal aspect that affords interactions between ionized or excited molecules and radicals within a track such that different chemical modifications may occur by multiple interactions in a molecule.

In addressing dose rate, the **Fig.20** describes the distribution of ionized regions (spurs) along a particle track and the proximity of these regions to DNA and to themselves. Independent of reactions and associated material responses it is useful nonetheless to consider simply the areal density of the particles (e.g. protons) themselves in a single bunch of short duration. Letting the areal track density in a cell be equal to the incident areal particle density of a single bunch (i.e. the longitudinal integration of volume density) one can estimate the areal density required for tracks to spatially overlap. Assuming a track area to be approximately equal to the estimated spur area of order 100 $nm^2$ the proton areal density required to exhibit a dose rate effect based on spatial overlap would exceed $10^{12}$ $cm^{-2}$. For 250 MeV protons this corresponds to a volume density near $6 \times 10^{10}$ $cm^{-3}$ (i.e.~ 10 nC $cm^{-3}$) representing a proton spacing of about 1.6 microns (and a temporal separation of 8 femtoseconds). Although this simplistic argument ignores spur and track-related lifetime issues the relevant timescale is clearly the particle bunch duration at the cell (~ nanoseconds).

Another viewpoint has also been suggested in which the time separation (and therefore longitudinal distance) between particles in a bunch is addressed by comparisons with characteristic response times of targeted cell material. In this model high dose rate effects (evidenced by enhanced RBE) might occur for bunch densities where the particle spacing is of order or less than the product of this characteristic response time and the particle speed **(108, 109)**. For example a cell response time (frequency) of 10 femtoseconds (100 THz) for 250 MeV protons is matched by a particle spacing of 2 microns and a single bunch density near $3 \times 10^{10}$ $cm^{-3}$ (a result similar to the preceding simplistic estimate based on spatial overlap). This cell response picture has been described in terms of a proton clustering effect at high bunch densities (notably for ultrashort bunch durations) that could exist near the source (laser target) to boost dose rates by as much as two orders of magnitude. These reported simulated results and the response rate hypothesis can be tested in cellular radiobiological studies.

A number of other papers dealing with the dose controlled irradiation of cancer cells with laser accelerated proton beams, laser accelerated electron beams and x-ray beams including the dosimetric uncertainties introduced by the broad energy spectra of laser driven particle beams have been published over the years **(100-103).**

*4.4 Oxygen Effect and Dose Rate*

We briefly consider here the cell and its environment, as opposed to the isolated cell of the preceding sections. The presence of oxygen is known to enhance cell lethality by a factor of about 3 for low LET radiation (**104**). This enhancement is probably due to the fact that DNA radicals readily react with oxygen to generate biologically relevant DNA damage (**105**). At ultra-high dose rates (~ $10^9$ Gy/second or more) for low LET radiation this oxygen effect is evident in hypoxic tissue, as several Gy dose levels deplete the oxygen level at such dose rates. Although oxygen is easily accessible to single monolayer cells, oxygen recovery to regions of diffusion-related hypoxia, which are often observed in a relatively small proportion of tumor tissues, can require significant time (several seconds for tissues having distances of more than 100µm from a tumor blood vessel) (**104**). This means that the efficiency of cell inactivation in a tumor needs to be optimized or improved with the mode of irradiation featuring breaks (brief time intervals with no irradiation) of adequate duration (~seconds) such that all cells in a tissue are accessible to oxygen.

We suggest that this important oxygen effect also needs to be studied with laser-driven moderate LET (proton) irradiation, since the dose required to deplete oxygen is unknown and could be different with low LET irradiation. In contrast we note that high LET irradiation exhibits no such oxygen effect. Lethality enhancement due to the presence of oxygen rapidly diminishes if the LET exceeds ~ 60 KeV/micron (**90**). This is possibly attributed to the decomposition of water with high ionization density or an effect of the increased clustering of lesions.

*4.5 Laser-driven radiation in radiobiological studies*

In 2012 the RBE of electrons and protons was addressed (**23, 106-108**) and the overall results show no significant differences in radiobiological response for *in vitro* cell experiments between pulsed laser-accelerated and conventional clinical electron beams. Much the same conclusion has been drawn in comparisons between protons from laser-driven and conventional accelerator sources.

At this point we also address the potential use of laser-driven radiation with ultra-high dose rates in future radiobiological studies. A basic cellular radiobiological agenda will confirm the usefulness of laser-accelerated ion irradiation with effects similar to those generated with conventional accelerators. Although radiobiological effects which originate from DNA damage are less likely to exhibit any ultra-high dose effects, various biological endpoints derived from non-DNA targets or multi-cellular phenomena may indeed reveal dose rate issues. However, we expect that, under any circumstances, the ionized/excited molecules or radicals generated at biologically relevant doses should interact with one another to cause ultra-high dose rate effects. The use of microbeams to reduce the spatial distance between the tracks could increase the possibility of such interactions. One of the promising directions using laser-driven radiation may also involve studies of very fast physiochemical processes that take place within the picosecond to nanosecond timescale after irradiation of biomolecules.

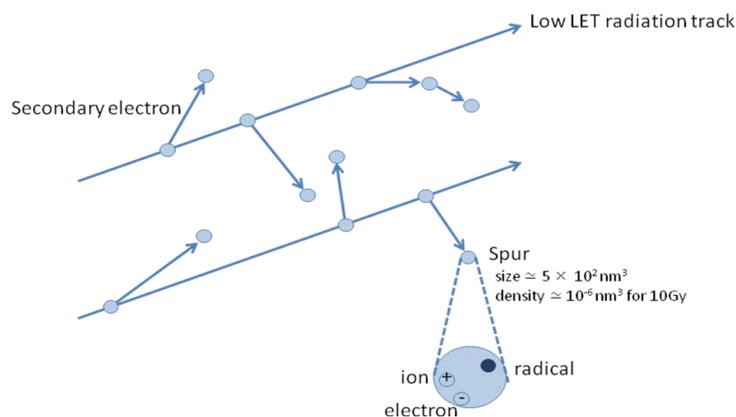

**Fig. 20.** Schematic representation of the tracks of LET radiation and spurs.

## 5. Real time Identification of healthy and cancerous tissue using Laser Desorption Ionization Mass Spectrometry

In the 1980-90s high intensity laser driven ionization and fragmentation mass spectrometry was developed for the ultra-sensitive detection particularly of molecules **(119)**. The applications were principally for the sensitive detection of explosives, drugs of abuse and environmental pollutants.

Most hydrocarbon and bio-molecules when irradiated by an intense laser beam fragment into moieties which are characteristic of that molecule but which usually contain hydrogen, carbon, oxygen, nitrogen and other atoms in certain proportions.

During the last few years Takats and his team **(120-124)** realised that this technology could also be used to distinguish between healthy and diseased cancerous tissue during the removal of malignant tumours which can potentially enable surgery with greater precision and reduce the likelihood of further surgery. In particular a neurosurgery challenge is the distinction of where a brain tumour ends and healthy tissue begins. This procedure could be carried out in real time.

In **Fig. 21** the fragmentation patterns of certain tissues are shown and it can be seen from B and C the ratios of certain masses around 700 amu are different for healthy and diseased tissue. The fragmentation patterns can be induced by either lasers or electrical high temperatures. The ratios of the principal components (the highest mass peaks) are measured and it is found that the variance of any specific tissue (healthy or diseased) is smaller than the variance between tissue types (healthy or diseased)

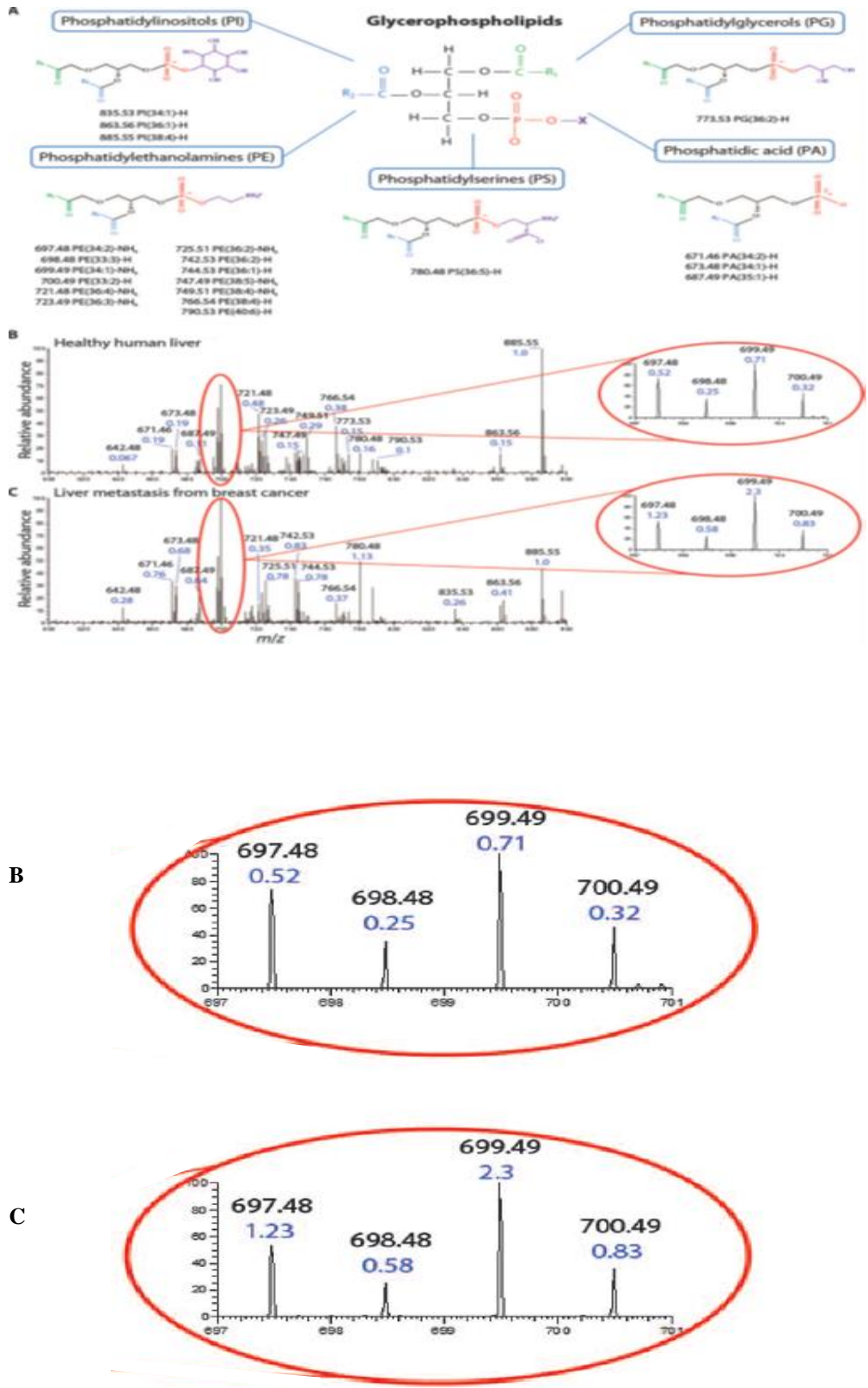

**Fig.21)** This figure shows the fragmentation pattern of human tissues both healthy and diseased. There are certain prominent mass peaks which are called principal components and it is found that the principal components of healthy and diseased tissues are sufficiently different to be used as a diagnostic technique. Reproduced by kind permission of Takats *et al* (122) and the publishers.

A tumour is shown in **Fig.22** below surrounded by healthy tissue

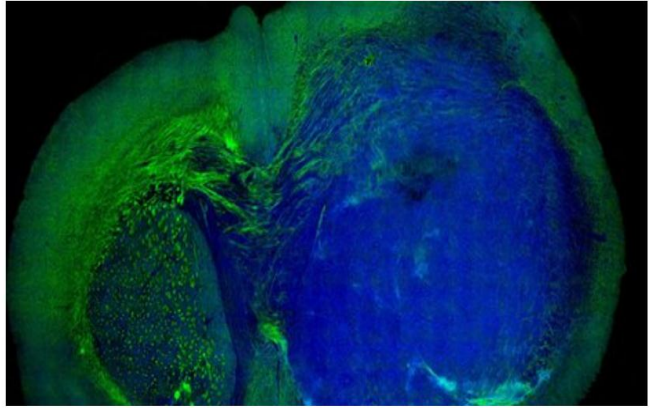

The edges of the tumour in blue are clear against the green healthy tissue

**Fig. 22** This figure shows a tumour (in blue) surrounded by healthy tissue (in green). Reproduced by kind permission of James Gallacher (BBC News) **(125)**

If we use high power lasers to produce particle beams for therapy of diseased tissues, we can simultaneously (& synchronously) use the laser beams to produce fragmentation patterns of the tissue to measure the precision of the beam therapy or surgery. The intrinsic capability of the laser to provide multiple synchronous beams for multiple related functions is one of its most valuable attributes.

## 6 Some notable sites (current and planned) exploring laser-acceleration and potential medical applications

Initial details of the research centres and the specifications of some of the lasers involved in laser driven therapy and applications are given in an ICFA Publication **(110)**

*a) Queens University Belfast Consortium*

UK activities aimed to develop laser-driven ion sources for medical therapy are currently carried out within the framework of the A-SAIL consortium, which involves researchers at Queen's University Belfast, University of Strathclyde, Imperial College and Rutherford Appleton Laboratory. This project, started in May 2013 aims to focus more closely on medically-related objectives. The main aim of the initiative is to unlock, within the next 6 years, the potential of the laser-driven ion approach as an alternative source for cancer therapy, advancing laser-ion sources to the point at which they can be a competitive alternative to conventional RF accelerators for medical therapy. The project aims to demonstrate controlled, all-optical acceleration of dense bunches of protons and other low-Z ion species in the 60-300 MeV/nucleon range of interest for therapy of deep-seated cancer. A number of different laser acceleration protocols will be studied to increase the energy of the ion beams e.g. Radiation Pressure Acceleration. Achieving this objective will require a coordinated effort involving development of new target media, understanding and controlling the physical processes of the relevant interaction regimes, and developing innovative solutions to known difficulties. A unique property of laser-driven ion beams is their ultrashort duration since ions are emitted in bursts of ps or fs duration at the source and their therapeutic use may result in dose rates up to 9 orders of magnitude higher than normally used in therapy. Dose rate can be a critical parameter affecting radiation biological responses. The biological effects of ions at such extremes are virtually unknown, and need to be carefully assessed. In parallel to the source development, the consortium is pursuing, by employing the unique capabilities of current and developing laser systems in terms of ion production and delivery, a programme of investigations of the dynamics of cellular response to ion irradiation and damage at these unprecedented dose rates. The first experiments exploring the biological effects on cells at dose rates $>10^9$ Gy/s have been carried out on QUB's TARANIS laser, and recently on the GEMINI laser at RAL **(108)**

*b) Laboratoire d'Optique Appliquée ~ LOA, Palaiseau, France*

As has been described in the text, at LOA, both electron and proton beams **(111)** produced by laser plasma accelerators are considered as potential candidates for radiotherapy. With respect to protontherapy, LOA hosts the SAPHIR facility. The corresponding project called SAPHIR involves academic (groups from CNRS and CEA), medical groups (from Institut Gustave Roussy and Institut Curie), and industrial partners (Amplitude Laser Technologie, Imagine Optics and Dosisoft). The installation of the 200 TW laser developed by Amplitude Technologie, at LOA is now finished to enable first experiments. It is expected for this phase of the project that proton beams with energies in the few tens of MeV will be produced, and that proton beams will be shaped to allow the first radiobiological studies. These results will be compared to theoretical and simulation predictions that will help in defining the laser and target performances required to reach proton beams of 150-200 MeV which is necessary for the therapy of deep seated tumours.

In addition the use of energetic electron beams for radiotherapy has been a major emphasis of the centre. The use of electron beams at moderate energies, 5 to 20 MeV range, for treating cancer and disease, is important because of the low penetration range in the patient. This is limited to surface tumours, i.e. tumours or disease located a few centimetres below the skin and with negligible dose to surrounding sensitive tissues. Since the range of 20 MeV electrons in water is about 10 cm while the range of 150 MeV electrons is more than 40 cm, higher energy electrons are potentially good candidates for improving cancer treatment of deep seated tumours. Moreover in contrast to present clinical electron beams, scattering in air of high-energy electrons is sufficiently small that electro-magnetic pencil beam scanning is possible. This could benefit intensity modulated treatment where high lateral resolution is necessary. It has been shown that the range of energy, the charge per bunch, and the repetition rate of the laser satisfy many of the medical requirements for this application **(84-88).** Monte Carlo calculations have been found to be in very good agreement with dose measurements. Treatment planning for laser accelerated high energy electrons has already been carried out for prostate irradiation and further studies are scheduled for other organs

*c) Ion Acceleration Program at BNL ATF and UCLA*

By expanding the spectral range of available laser drivers to the mid-IR region, the 9-10μm carbon dioxide ($CO_2$) lasers offer new opportunities to explore ion acceleration.

Several advantages of the long-wavelength radiation, i.e. stronger ponderomotive electron heating and reduced critical plasma density, facilitate access to the Shock Wave Acceleration (SWA) regime from gas jets. Attractive features of this mechanism are: monoenergetic ion energy spectra, near linear ion energy scaling with the laser intensity, and targets of pure composition that are better suited for high repetition rate operation. These features have been experimentally demonstrated with picosecond $CO_2$ lasers operated at UCLA's Neptune laboratory **(48)** where high-brightness pure proton beams of 3-20 MeV energy with 4-10 percent energy spread have been generated with supersonic hydrogen jets and at the Accelerator Test Facility (ATF) at the Brookhaven National Laboratory (BNL) **(49)**.

The BNL's ATF is a DOE user facility open for international collaborations. The SWA experiment at ATF has been conducted in close partnership with Imperial College (London) and Stony Brook State University (New York). Further progress of this collaborative effort at ATF towards ion beams with kinetic energies suitable for medical applications relies on the current $CO_2$ laser upgrade to the 50-100 TW peak power range. Preliminary work suggests that protons with energies approaching 100 MeV in a quasi-monoenergetic peak may be reached if $a_0$ values greater than 6 can be achieved.**(112).** This next-generation ultra-fast (pulse duration near 10 cycles) $CO_2$ laser will integrate several innovative design features such as pulse stretching and compression that has not been previously implemented with molecular gas lasers, isotopic amplifiers, and nonlinear self-chirping and filtering.

When this system is completed, one of the long term campaigns will be aimed at proton (and other ion) acceleration by exploring collisionless shock techniques. The experimental work will be supported by PIC code simulations. This is to be a joint undertaking led by UCLA and BNL. The upgrade underway at ATF is a three-year plan that includes dedicated experimental space for applying laser-produced ion beams to user experiments with biological samples and living tissue. The Stony Brook University Department of Clinical Oncology will join the effort during this user stage.

*d) OncoRay - National Center for Radiation Research in Oncology, Dresden*

A long-term aim of OncoRay with its three operating partners (Technische Universität Dresden, University Hospital Carl Gustav Carus Dresden and Helmholtz-Zentrum Dresden-Rossendorf) is the development of laser-based ion-beam radiation therapy technology for clinical treatment of cancer patients. The necessary translational research from basics to clinical application and close multidisciplinary cooperation of physicists, engineers, biologists and physicians is supported by the German government (joint research project onCOOPtics pursued in cooperation with Ultraoptics Jena, second funding period 2012-2017), the Free State of Saxony and the European Regional Development Fund.

During the last few years, laser-based irradiation technology has developed to such an extent that cell samples and small animals can be irradiated by a fixed beam line as is already done in ongoing systematic radiobiological experiments. Further translation towards clinical application is focusing on precise and efficient dose delivery to large tumours in patients and involves two complementary topics:

(i) Development of energy-efficient high-power laser systems and laser targets for the routine generation of clinically relevant ion beams.
(ii) Development of medical beam delivery units for laser-driven ion pulses including compact rotating gantries based on pulsed-power magnets, precise 3D dose monitoring in real-time and efficient treatment planning strategies.

HZDR operates two high power laser laboratories with dedicated infrastructure for the investigation of laser particle sources. The first hosts a Petawatt Ti:Sapphire ultrashort-pulse laser system (30 J in 30 fs) embedded into an accelerator laboratory. The second serves for the development of a fully diode-laser pumped energy-efficient solid state Petawatt laser that may serve as a prototype for clinical applications.

A new research building was constructed on the campus of the university at a Dresden hospital. It is connected to the department of radio-oncology and hosts a conventional proton therapy facility consisting of a cyclotron, a beam line with an energy selection system and one treatment room with a rotating gantry (at present under commissioning, with the start of patient treatment in 2014). In addition, research facilities for cell and animal experiments as well as a Petawatt laser laboratory are available. Moreover, an experimental bunker allows the use of both the laser-based experimental proton beams and the reference beam from the cyclotron.

*e) Laser-acceleration Studies at JAEA in Japan*

Electron and proton acceleration with lasers has been an active area of investigation at the Kansai Photon Science Institute (KPSI) of JAEA for several years. Researchers at KPSI explore candidate laser-driven proton sources and the potential for various applications. For example KPSI has demonstrated focusing and spectral selectivity of 1.9 MeV protons at 1 Hz **(73)**. The RBE of 2.3 MeV laser-accelerated protons irradiating invitro cells of a human salivary gland tumour has been measured with 0.2 Gy single bunch doses delivered at 1 Hz **(25,113)**. In recent laser experiments protons have been accelerated to energies as high as ~ 43 MeV (cut-off value) with an on-target laser intensity (power) of $10^{21}$ W/cm$^2$ (200 TW) and an intensity contrast at the $10^{-10}$ level **(136)**.

Experiments have been supported historically by strong laser-plasma simulation efforts (see for example **44, 45-51**). Furthermore, in collaboration with the University of Bologna, the University of Milano and INFN the use of multi-layered targets as sources combined with candidate charge collection/transport optics have been investigated in simulations of a hybrid ILDIAS (i.e. HILDIAS) in which final acceleration to a desired operating proton energy is accomplished using a conventional compact side-coupled RF linac) **(137)**.

Currently, the J-KAREN laser at KPSI is being upgraded to higher pulse power and repetition-rate capability (with the upgraded result to be renamed as J-KAREN-P). J-KAREN-P will deliver an on-target peak power of ~ PW at 0.1 Hz with improved focusing and intensity contrast. This will enable extended secondary source studies. With higher intensity on target KPSI aims to make notable progress toward reaching ~ 200 MeV proton energies as well as KeV x-rays.

*f) The Munich Centre for Advanced Photonics (MAP)*

The Munich-Centre for Advanced Photonics (MAP, http://www.munichphotonics.de) is a research cluster funded by the German government since 2000. MAP draws on recent developments in high-power laser technology and emerging secondary sources of laser-driven X-rays and laser-driven particle beams for (i) imaging and early detection of small tumours (before they can form metastases) and (ii) high-precision image-guided local radiation therapy. As previously discussed, laser-based particle acceleration (in particular of protons and carbon ions) offers the potential of compact and cost-efficient sources that could replace conventional cyclotrons or synchrotrons. This means that more cancer patients (both with small and advanced tumours) could benefit from particle beam radiotherapy and particularly in countries where RF conventional accelerator based technology is too expensive.

Researchers at MAP are currently investigating the required physical, technological and biological basis for laser-based radiation therapy with proton or carbon ion beams. They focus on these issues both from a fundamental physics point of view as well as with respect to applied medical physics and radiobiology with the long term goal of developing a laser-based particle therapy unit. In particular, the ion acceleration process itself is studied using ultra-thin diamond-like carbon foils as targets, and a laser-driven biomedical beamline for cellular radiation biology and dosimetry has been constructed.

As soon as higher proton energies (250 MeV) will be available, radiobiological investigations will shift to pre-clinical studies in mice and finally towards clinically relevant setups. Appropriate methods for beam delivery including lateral and axial beam shaping for highly pulsed beams and the corresponding treatment planning strategies are being developed in order to design clinical treatment units for laser-driven particle beams. These activities are pursued by MAP and at the new facility "Centre for Advanced Laser Applications" (CALA), a dedicated new building on the Garching research campus (near Munich). The long-term goal is development of an integrated, laser-based system for high-resolution phase-contrast X-ray imaging and particle therapy of tumours.

*g) ELIMED Prague*

The ELIMED (ELI MEDical applications) research project has been launched by INFN-LNS and ASCR-FZU researchers within the pan-European ELI-Beamlines facility framework. The ELIMED project aims to demonstrate potential multidisciplinary user applications of laser-accelerated proton beams. In this picture eye melanoma (uveal melanoma which is normally treated with 62 MeV proton beams produced by standard accelerators) will be considered as a model case to demonstrate a potential future use of laser-driven protons in hadrontherapy; especially because of the limited constraints on proton energy and irradiation geometry for this particular tumour treatment.

Several challenges, starting from laser-target interaction and beam transport development up to dosimetry and radiobiology, need to be overcome in order to reach the ELIMED final goals. A crucial role will be played by the final design and realization of a transport beamline capable of providing ion beams with proper characteristics in terms of the energy spectrum and angular distribution which will allow performing dosimetric tests and biological cell irradiation. A preliminary prototype transport beamline has been designed and some transport elements are under construction for first experimental tests at the TARANIS laser system.

A wide international collaboration among specialists of different disciplines like physics, biology, chemistry, medicine and physicians coming from Europe, Japan, and the US is growing around the ELIMED project. The aim is to create a network of specialists proposing conceptual designs, and technical and scientific solutions likely to be implemented at the ELI Beamlines facility.

*h) Selcuk/Strathclyde Consortium Konya, Turkey*

Initially the existing lasers in Konya will be applied to studying laser ablation fragmentation patterns of both healthy and cancerous tissue from both animal and human tissue (**120-124**). This research will afford determination of tumour edges with greater precision. Along with precision surgical procedures this will reduce the incidence of re-growth of tumours that can necessitate future surgery. Selcuk University and the University of Strathclyde are part of an international collaboration (TILADIA) to carry out tissue analysis by direct laser desorption mass spectrometry for the stratified diagnostics of cancer patients. This is headed by Dr. Takats (Imperial College).

Following research with tissue analysis, laser-acceleration of protons to ~ 70 MeV energy will be pursued to enable irradiation of small animal tumours. This program is expected to take about two years to complete. The potential for direct irradiation of tumours inside the body by high power lasers via optical fibres will also be investigated (i.e. use the tumour as a target to produce the proton beams). This has the added attraction of requiring less energetic protons to irradiate deep seated tumours. The longer term vision is to site an international centre at the Selcuk University Hospital to develop laser driven proton and heavy ion beams for cancer therapy.

## 7. Comparative Radiotherapy Costs

We stated in the preamble that there are some 39 conventional accelerator centres around the world used for proton and heavy ion cancer therapy. The sites in Europe are shown in **Fig. 23**

It has been mentioned (**3**) that particle therapy is potentially a more effective therapeutic procedure than photon therapy. How costly is particle therapy? Andrea Peeters et al (**114**) have estimated the capital costs for a) combined proton /carbon facility, b) proton only facility and c) photon facility to be a) 139 million €, b) 95 million € and c) 23 million €.

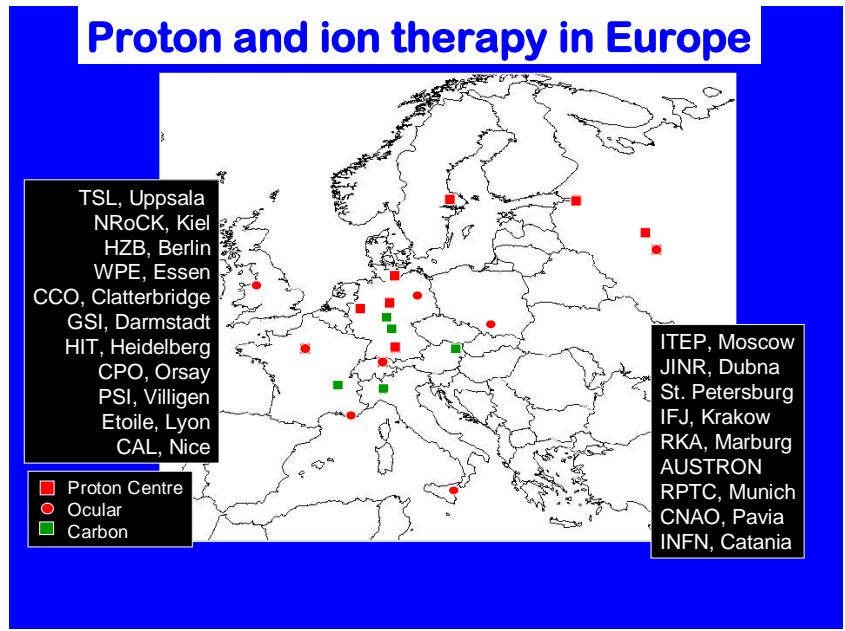

**Fig. 23** Proton and ion therapy centres in Europe   Reproduced by kind permission of the Dr Andrzej Kacperek ,Clatterbridge Centre,Liverpool

The relative cost of treatment can also depend on where the tumour is located. The high cost of accelerator hadron treatment is evident when one considers the size of a typical hadron therapy facility (as shown for example in **Fig 24**). The cost figures presented in Peeters et al are very similar to those cited in **(3).**

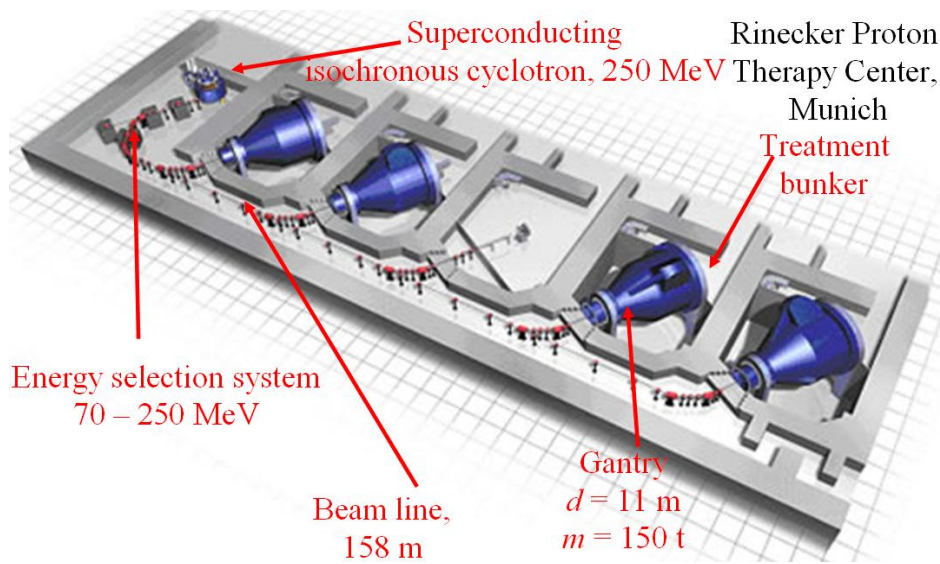

**Fig. 24** The Rinecker Proton Therapy Center in Munich.  The large capital costs can be mitigated if a number of treatment areas are provided which results in more patients being treated.  It is important to point out that in the future costs could be further reduced if compact cyclotrons are used.

Two very important systematic literature reviews of the clinical cost-effectiveness of conventional accelerator driven hadron therapy were carried out in 2007 and 2012 **(12,115)**. The conclusion of the first survey was "existing data do not suggest that the rapid expansion of hadron therapy (HT) as a major modality would be appropriate. Further research into clinical and cost-effectiveness of HT is needed." In the second survey some 5 years later, this conclusion was slightly tempered with the "aim of building enough proton therapy facilities and run clinical trials to draw firm conclusions about the clinical effectiveness of charged particle therapy to best current practice".

The cost of establishing a laser-based therapy centre can potentially be much cheaper although one must be cautious here because all essential components must be considered and no centre has yet been built. Significant cost reduction in the laser case can be possible if a) future high power lasers become much more compact and can therefore be housed in a smaller building b) radiation shielding can be significantly reduced and c) if an optical gantry were possible that could be lighter and smaller than that associated with a conventional accelerator-based system. Costs for a laser-driven facility have been estimated by a number of authors to be between 10-20 million € **(75,114,115)**. The Thales laser company recently developed the Bella laser (at a cost near $11 million) which is used at the Lawrence Berkeley National Laboratory. With petawatt peak power capability it has potential for accelerating protons to energies suitable for radiotherapy **(116).**

Although most developing countries would find the costs required for a conventional accelerator-based hadron therapy centre to be prohibitive they might find a laser-based system to be more affordable given national budgets. These countries could potentially afford 5-10 million € to initiate a laser-driven beam centre that can guide future innovation. Such startup laser systems may not be able to provide 250 MeV protons for deep tumour therapy but they can establish the necessary foundation and suitably comprehensive thinking.

Finally it should be emphasised that the center costs mentioned in this section are limited to the cyclotron and treatment rooms and for the basic laser and a single treatment room. Like-for-like comparisons have not been made rigorously for the two approaches to proton therapy.

## 8. Discussion and Conclusions

Before discussing in detail where we are with laser driven ion beams for oncology we comment briefly on possible controversies concerning the effectiveness of ion beam radiotherapy in general (i.e. with conventional accelerator driven proton and heavy ion beams). Two very important papers addressing this ongoing debate have recently been published **(126,128).** There is no question that charged particle radiation therapy offers significant advantages in physical dose distributions (Bragg Peak – **Fig.1),** however the authors of these two papers are questioning whether better physical dose distribution necessarily results in overall clinical benefit. The conclusions of these papers strongly suggest that randomised trials would remain the preferred goal to test the effectiveness of proton and heavy ion therapy. In addition, establishing a European Hadron Therapy Register (or better an International Register) to make better use of the data being generated by the many existing hadron therapy centres would be greatly beneficial. A further controversial article was published in the Biology and Medicine Journal ( entitled "Is spending on proton beam therapy for cancer going too far too fast" **(128).**

However in answer to this paper a group of UK oncologists claimed the report misrepresented the approach being used in the UK **(129).** Epstein claims that despite huge amounts of money being put into building accelerators for proton therapy the first randomised trial is just beginning and won't be completed for seven years.

In this work the essence of these considerations for the future is the development of reliable, controlled laser-driven sources (photons and particles) for medical application. Since we must wait seven years for completion of the first randomised trial that is testing the efficacy of conventional accelerator driven proton therapy, it is relevant to speculate about the anticipated state of laser-driven hadron therapy in the future. A number of articles dealing with the prospects for and progress towards laser proton therapy have been published especially in the last year dealing with this question **(117,129-133,)** with estimated time frames of 4-10 years or longer being suggested for implementing laser driven ion beam radiotherapy (shorter time estimates apply to ocular tumours for which required proton energies are ~ 70 MeV).

With appropriate targetry, contrast and focusing PW level lasers will likely accelerate proton and ion beams to kinetic energies that are relevant to radiotherapy of deep seated tumours. Over the next few years these systems will likely continue to be very large and very expensive. Thus to deliver the vision of a compact, affordable ILDIAS much "thinking out of the box" could be necessary to deliver better targets, better lasers and better beam delivery. In the broader context of fully functional integrated laser-driven systems (ILDIAS and ILDEAS) it is clear that considerable progress must still be made if we hope to realize laser-driven accelerators with medically relevant beam parameters and performance levels suitable for clinical usage. Meanwhile, the path toward laser-driven energetic particle beams of medical quality will no doubt also enable other important 'en route' applications that can serve to highlight ongoing progress and multiple laser-driven capabilities.

Although many researchers at high power laser laboratories around the world make reference enthusiastically to laser-driven proton radiotherapy, time on these lasers to develop the necessary clinical requirements can be quite limited and such applications are typically treated as a lower priority. There is little doubt that realizing concrete progress toward the laser-driven radiotherapy vision requires a laser centre that is dedicated to laser-driven accelerator development for radiotherapy and a variety of other applications. A dedicated centre for laser-driven particle beam lines and other secondary sources such as x-ray beams will highlight key unique features of laser-driven secondary sources. A centre like this would also need a laser-driven radiobiology program, biomedical technology program to support such applications and a globally coordinated preclinical science agenda. Throughout, the centre should remain adaptive to new developments. An overall programmatic strategy that is built around key intermediate milestone applications (not necessarily medical) is strongly recommended.

In conclusion we recognise that the state-the-art for ion beam radiotherapy (IBRT) is not static. Therefore comparisons with the laser-case (and with other cases) can be dynamic. For example, in parallel with progress made in the laser-case, commercial development of compact cyclotrons (notably superconducting synchrocyclotrons) and associated gantries is ongoing. A good example is the recent developments led by IBA to commercially provide a KHz proton source with nanoampere current at a kinetic energy up to 250 MeV **(134,135).** As with the optical gantry for the laser case, embedding an adequately compact cyclotron source into a single gantry is being considered. So, evolution of the cost/size comparison is not attributed solely to laser system development. It is clear that new source development for medical applications must be monitored continually and comprehensively.

With lasers in mind, it is therefore critical to identify noteworthy features of laser-driven particle acceleration that can also be unique. The potential for spectral (particle energy) control in the delivery of therapeutic treatment, for sophisticated targetry (for example, that can also function as an ion optic) and for particle bunches of high charge, short duration and very low emittance are some examples. The flexibility that a laser-driven source can afford is also noteworthy. This includes the potential for multiple particle beams that can include different ion species as well as an intrinsic synchronized multiple beam capability that can feature secondary (laser-driven) x-ray sources and diagnostic laser probes to accompany energetic particle beams. Of course, the requisite laser-plasma engineering needed to realize such capabilities is a great technical challenge that can be addressed at a centre that is dedicated to ILDIAS development. As the relevant laser, laser-plasma and target science and technologies mature we remain optimistic that cost/size comparisons will become increasingly favourable and the therapy niche increasingly clear.

## Acknowledgements

The authors take great pleasure to acknowledge the assistance provided by the following colleagues Toshi Tajima, Victor Malka, Marco Borghesi, Jorg Pawelke, Farhat Beg, Mark Sheenan, Mark Lodge, Kristofer Kainz, Jorg Schreiber, Jan Wilkens, Markus Roth, Danielle Kelly, Bleddyn Jones.

K.W.D.L. would like to thank the University of Selcuk, Konya, Turkey particularly Prof Hamdi Kilic and the whole medical and surgical team for huge support and encouragement including access to complicated surgery which could benefit from the science covered in this review.